\documentclass[prl,showpacs,twocolumn,superscriptaddress]{revtex4}

\usepackage{subfigure}
\usepackage{wrapfig}
\usepackage{textcomp}
\usepackage{graphicx}
\usepackage{amssymb}
\usepackage{amsmath}
\usepackage{bm}
\usepackage{amsmath, amsthm, amssymb}

\begin{document}
%\definecolor{darkgreen}{rgb}{0,0.5,0}

\newcommand{\bmm}{{\bf m}}
\newcommand{\bb}{{\bf b}}
\newcommand{\ba}{{\bf a}}
\newcommand{\bj}{{\bf j}}
\newcommand{\bA}{{\bf A}}

\title{Fractional topological phases and broken time reversal symmetry in strained graphene}

\author{Pouyan Ghaemi}
\email{pouyan@berkeley.edu}

\affiliation{Department of Physics, University of Illinois, Urbana, IL 61801}

\affiliation{Department of Physics, University of California, Berkeley, California 94720, USA}

\affiliation{Materials Sciences Division, Lawrence Berkeley National Laboratory, Berkeley, CA 94720}

\author{J\'er\^ome Cayssol}
\affiliation{Department of Physics, University of California, Berkeley, California 94720, USA}
\affiliation{Univ. Bordeaux and CNRS, LOMA, UMR 5798, F-33400 Talence, France}
\affiliation{Max-Planck-Institut f\"ur Physik Komplexer Systeme, N\"othnitzer Str. 38, 01187 Dresden, Germany}

\author{Donna N. Sheng}
\affiliation{Department of Physics and Astronomy, California State University, Northridge, California 91330, USA}

\author{Ashvin Vishwanath}
\affiliation{Department of Physics, University of California, Berkeley, California 94720, USA}
\affiliation{Materials Sciences Division, Lawrence Berkeley National Laboratory, Berkeley, CA 94720}

\begin{abstract}
We show that strained or deformed honeycomb lattices are promising platforms to realize fractional topological quantum states in the absence of any magnetic field. The strained induced pseudo magnetic fields are oppositely oriented in the two valleys \cite{mpr,guinea2010,low2010} and can be as large as 60-300 Tesla as reported in recent experiments \cite{levy2010,gomes2011}. For strained graphene at neutrality, a spin or a  valley polarized state is predicted depending on the value of the onsite Coulomb interaction. At fractional filling, the unscreened Coulomb interaction leads to a valley polarized Fractional Quantum Hall liquid which spontaneously breaks time reversal symmetry.  Motivated by artificial graphene systems \cite{gomes2011, West, Sengstock,tarruell2011}, we consider tuning the short range part of interactions, and demonstrate that exotic valley symmetric states, including a valley Fractional Topological Insulator and a spin triplet superconductor, can be stabilized by such interaction engineering. 
\end{abstract}


\maketitle

\bigskip

Fractional Quantum Hall (FQH) phases are macroscopic scale manifestations of quantum phenomena with unique features including the fractional charge and statistics (abelian or nonabelian) of elementary excitations. This topological order originates from the strong Coulomb interactions between electrons moving in a partially filled Landau level induced by a strong magnetic field. Recently Chern insulator models with a nontrivial flat band \cite{tang2011,sun2011,venderbos2011} were also shown to exhibit topological order in the absence of any magnetic field \cite{neupert2011,donna2011a,donna2011,xgwen2,regnault2011,xlqi2011}. Those so-called Fractional Chern Insulators (FCIs) explicitely break time-reversal symmetry $\mathcal T$ as did the original Haldane model \cite{haldane1988}. In contrast, fractional topological insulators (FTIs) \cite{levin2009,maciejko2010,Swingle2011,neupert2} can be naively thought of as two copies of time-reversed Laughlin FQH states, thereby obeying time reversal symmetry $\mathcal T$. In spite of few proposals \cite{Ran2011,Vandenbrink2011}, the experimental implementation of FCIs and FTIs remains very challenging. 

Motivated by recent experimental advances \cite{levy2010,gomes2011}, we introduce another route towards fractional topological phases making use of the gauge fields that can be generated in a deformed honeycomb lattice \cite{mpr,guinea2010,low2010}. The associated effective magnetic fields are opposite in the two different valleys and therefore they do not break the time reversal symmetry $\mathcal T$ \cite{mpr}. Indeed, a scanning tunneling spectroscopy study \cite{levy2010} confirmed that straining graphene could yield flat Pseudo Landau Levels (PLLs) \cite{guinea2010,low2010} with effective fields as high as $300$ T in each valley. Most recently by designing a molecular honeycomb grid of carbon monoxide molecules on top of a copper surface, Gomes at al. \cite{gomes2011} were able to observe the linear dispersion of Dirac fermions in graphene, and furthermore to generate nearly uniform pseudo-magnetic fields as high as $60$ T by deforming this grid \cite{gomes2011}. Finally other realizations of artifical graphene systems, in patterned GaAs quantum wells \cite{West} or with cold atoms trapped in hexagonal optical lattices \cite{Sengstock,tarruell2011}, also provide experimental platforms to create strong valley-dependent effective magnetic fields.

%including the valley FTI state and flat band superconductivity \cite{uchoa2007,kopnin2011,kiesel2011}. FQH phases and superconductivity are two of the most significant phenomena of modern condensed matter physics which realize qualitatively different types of quantum order at macroscopic scale. Generally it is very difficult to observe those two phenomena in a single system because strong magnetic fields are mandatory for the usual FQH effect and harmful to most superconductors.

In this Letter, we first consider real graphene under strong pseudo-magnetic fields generated by a mechanical strain. We investigate the interaction-driven phases in the $n=0$ PLL using mean field and numerical exact-diagonalization. The unscreened Coulomb interaction stabilizes a valley polarized Laughlin liquid at filling $2/3$ of the $n=0$ PLL. This states breaks spontaneously time reversal symmetry and is characterized by a finite charge Hall effect. At the neutrality point, we predict that strained graphene have either spin polarized or a valley polarized state, depending on the strength of the on-site interactions, with current estimates \cite{wehling2011} favoring the former state. Second we have investigated what type of interactions could destabilize the valley-polarized 2/3 state towards a valley-symmetric (time-reversal invariant) FTI. It turns out that the 2/3 state is rather robust for realistic interactions. Neverthess attractive local corrections to the Coulomb interaction can stabilize this valley FTI, which is a FTI where the valley plays the role of spin. Finally further increase of the attractive interactions leads to a spin triplet superconductor. Since the reported effective magnetic field strengths are around $300$ T \cite{levy2010} or $60$ T \cite{gomes2011}, the predicted phases might conceivably be realized with larger energy gaps than in FQH states under a real magnetic field.

{\bf Model.} The noninteracting part of our model has been proposed by Guinea et al. in order to realize PPLs in strain graphene under zero magnetic field \cite{guinea2010}. The corresponding tight-binding Hamiltonian reads
\begin{equation}
H_0 =\sum_{\bold{r}_{i}} \sum_{a=1,2,3} (t+\delta t_{a} (\bold{r}_{i})) ( a^\dagger (\bold{r}_{i}) b (\bold{r}_{i} + \mbox{\boldmath$\delta$}_a) + h.c. ),
\label{hamiltonian1}
\end{equation}
where $\delta t_{a} (\bold{r}_{i})$ is the strain-induced variation of the nearest neighbor hopping amplitude (with respect to the unperturbed value $t \simeq 2.7$ eV) between A-site at $\bold{r}_{i} $ and B-site at $\bold{r}_{i} + \mbox{\boldmath$\delta$}_a$ of the bipartite honeycomb lattice \cite{mpr}. The smooth deformation field $\delta t_{a} (\bold{r}_{i})$ is chosen in such a way to produce a nearly uniform magnetic field with a valley-dependent sign \cite{guinea2010,VaeziPreprint}. The  valley dependent vector potential $\textbf{A}_\xi(\textbf{r})=\xi \sum_{a=1,2,3} \delta t_{a}(\textbf{r})e^{i \textbf{K}.\mbox{\boldmath$\delta$}_{a}}$ minimally couples to linearly dispersing low energy excitations near the Dirac points located at momenta $\xi {\bf K}$ with ${\bf K}=(4 \pi/3 \sqrt{3}a_0) {\bf e_x}$ and $\xi=\pm 1$, $a_0$ being the carbon-carbon bond length \cite{mpr}. The uniform pseudomagnetic field induces a pseudo Landau level (PLL) electronic structure $E_n=\xi \sqrt{2 e \hbar
  v_F^2 B |n|}$, where $n$ is the relative integer labelling the nearly flat levels (see supplementary). Beside the macroscopic orbital degeneracy, each of those PLLs has a four-fold degeneracy associated with the spin and valley isospin degrees of freedom. In contrast to the full $SU(4)$  symmetry of graphene in an external real magnetic field \cite{alicea2006, goerbig2006}, the internal symmetry of strained graphene is $SU(2)$ for the spin and only Z$_2$ for the valley degree of freedom.

In this work, we study interaction effects within the partially filled zero-energy flat band ($n=0$ PLL) created by strain. We consider the following interaction Hamiltonian on the honeycomb lattice:

\begin{align}
\label{hamiltonian2}
H_{int} &=\sum_{\bold{r}_{i} \not= \bold{r}_{j}} V({\bf r}_i-{\bf r}_j)n({\bf r}_i)n({\bf r}_j) + U_0 \sum_{\bold{r}_{i}} n({\bf r}_i) n({\bf r}_i) \\ \nonumber
 &+ U_{nnn} \sum_{\langle  \bold{r}_{i},\bold{r}_{j} \rangle } n({\bf r}_i)n({\bf r}_j),
\end{align}
where $V({\bf r}_i-{\bf r}_j)=e^2/4 \pi \epsilon|\bf r_i - \bf r_j|$ denotes the bare Coulomb potential, $n({\bf r}_i)$ the fermion number operator on site ${\bf r}_i$, and $\langle  \bold{r}_{i},\bold{r}_{j} \rangle$ represents summation over all pairs of next-nearest-neighboring (NNN) sites. The bare Coulomb interaction is the dominant interaction due to the poor screening in neutral graphene. Nevertheless we also allow arbitrary modification of the short-distance part of the Coulomb interaction by adding local on-site and NNN interactions with respective strengths $U_0$ and $U_{nnn}$. The nearest neighbor interaction is not effective in presence of strong pseudo-magnetic field because in the zero energy PLL the noninteracting wave functions are localized on a single sublattice (see Supplementary).  Interesting proposals for altering short ranged interactions using substrates with momentum dependent dielectric susceptibility has been discussed \cite{abanin2011}. Unfortunately the actual values of $U_0$ and $U_{nnn}$ are not known in strained graphene although first principles calculations yield total on-site coupling $U_0 = 9.3$ eV and a small deviation $U_{nnn} \simeq -0.04 e^2/4 \pi \epsilon a_{0}$ of the NNN coupling from its bare Coulomb value in freestanding (and unstrained) graphene in zero magnetic field \cite{wehling2011}.

\begin{figure}[t!]
\vspace{-2.cm}
\hspace{-1cm}
\subfigure{\includegraphics[width=5.2cm,height=6.4cm]{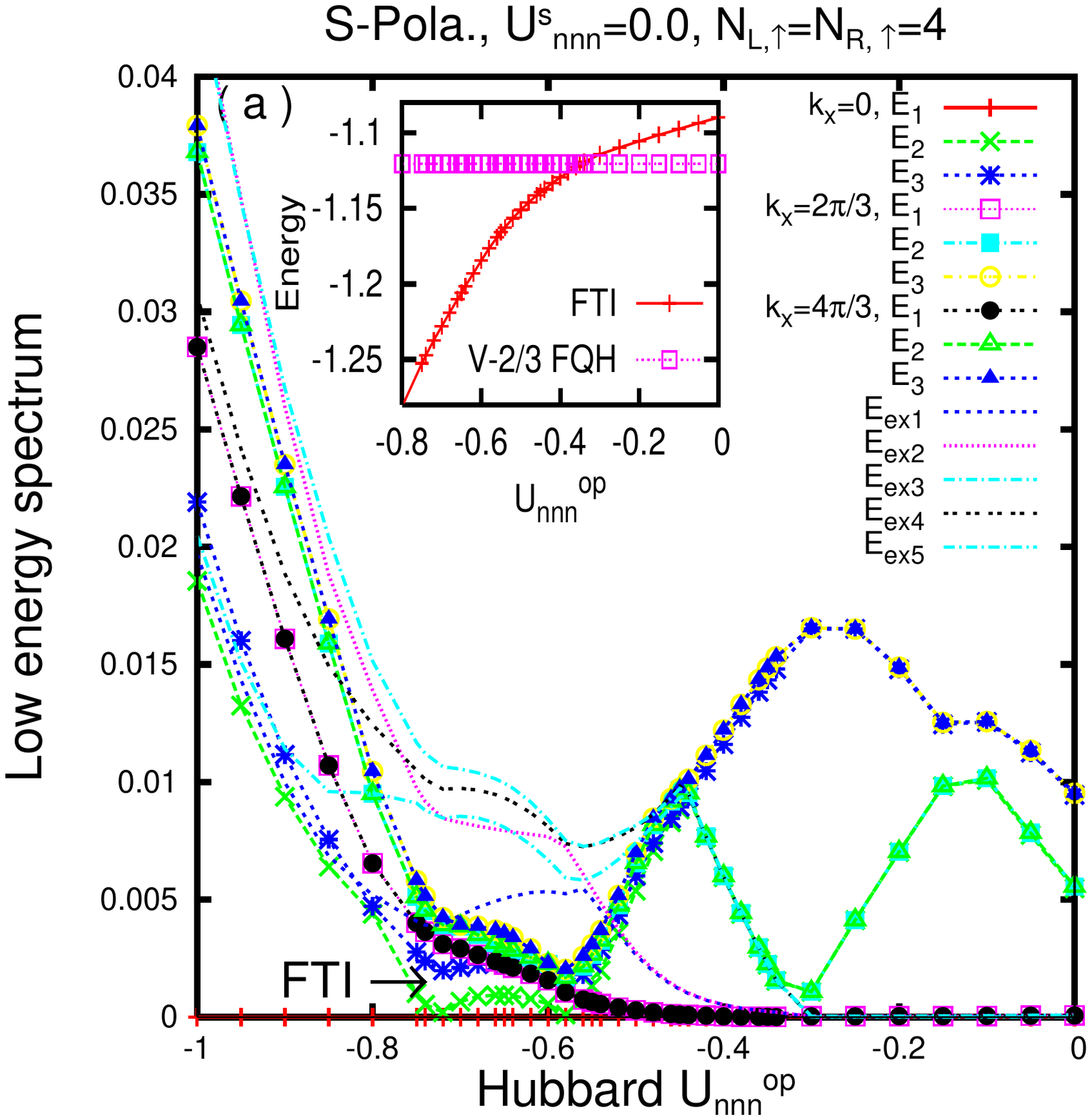}}
\hspace{-1cm}
\subfigure[]{\includegraphics[width=5.cm,height=6.2cm]{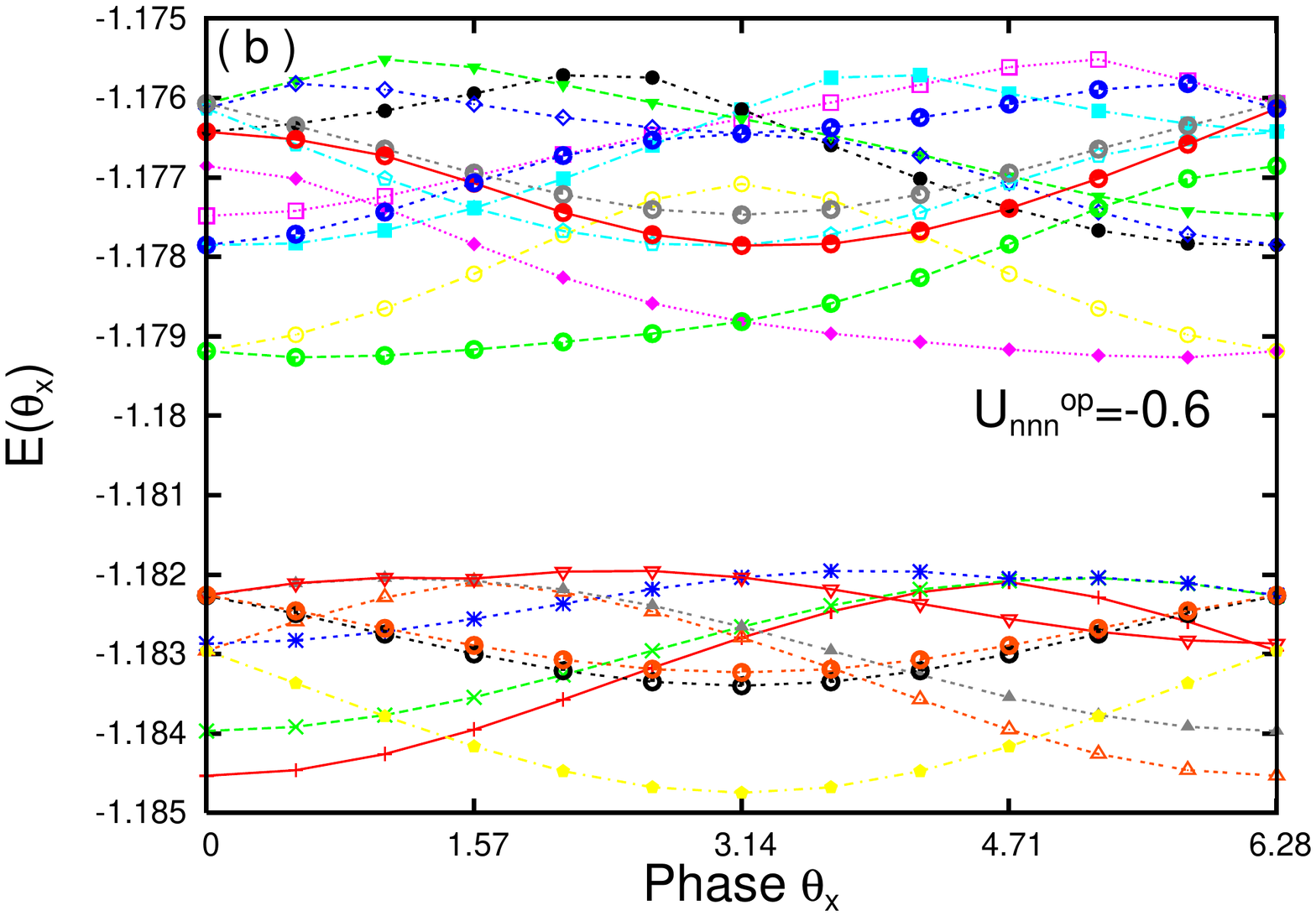}}
\vspace{-0cm}
\subfigure{\includegraphics[width=7.4cm,height=3.2cm]{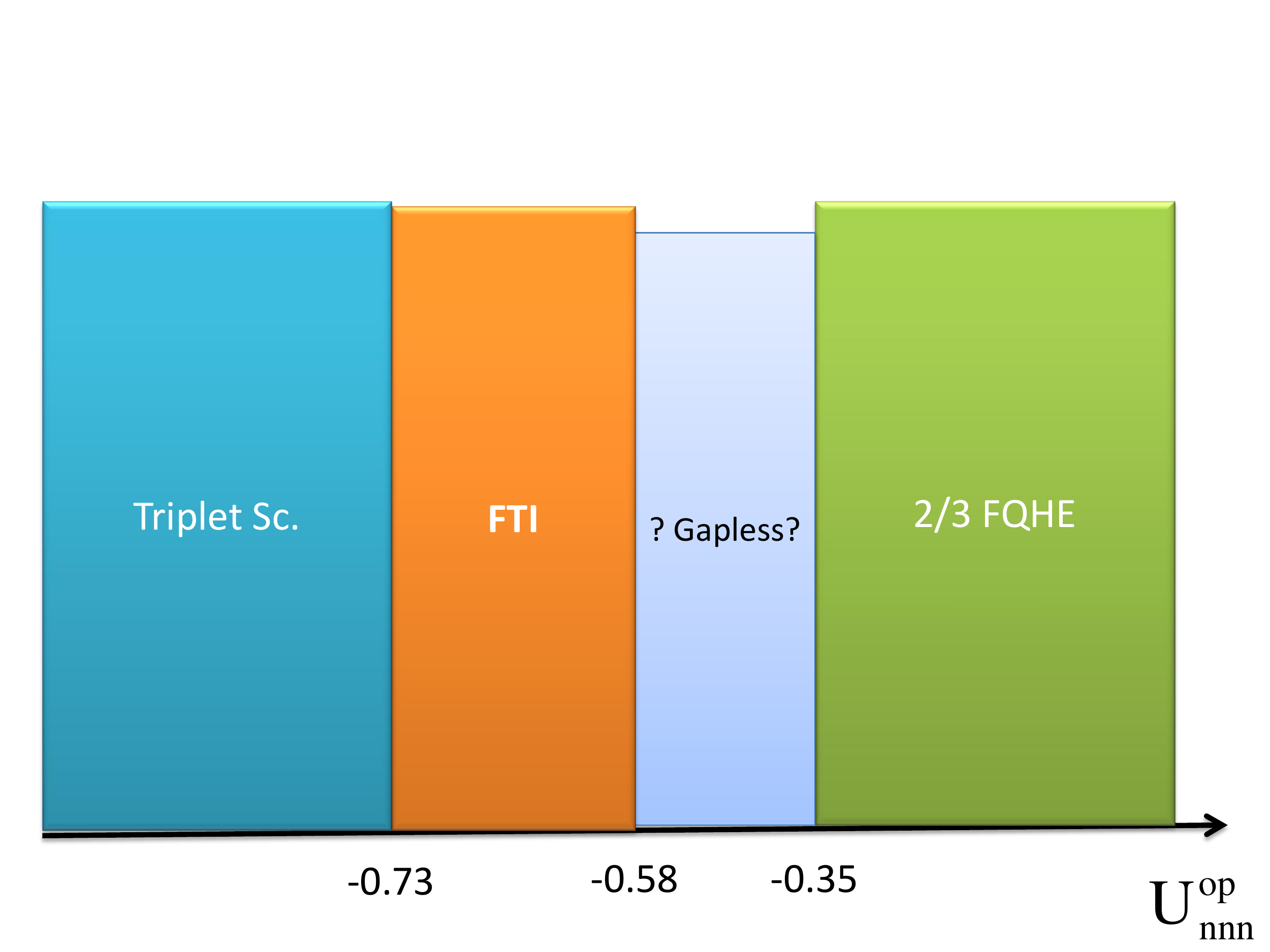}}
\vspace{-0.5cm}
\caption{(Color online). {\bf The $n=0$ PLL at fractional filling factor $\nu=-2+2/3$: spinless electrons.} {\it Upper panel left:} Low energy spectrum  as a function of the next-nearest-neighboring coupling $U_{nnn}^{op}$ between opposite valleys (deviation from the pure Coulomb value). 
In the region $-0.73<U_{nnn}^{op}<-0.58$, the nine lowest energy states become close
together and almost degenerated, thereby forming the groundstate manifold (GSM) of the valley FTI. The inset shows the 
groundstate energies of the valley polarized (V-2/3 FQH) and FTI states.
{\it Upper panel right:} The boundary phase dependence and the robust gap between the GSM and
higher energy states for $U_{nnn}^{op}=-0.6$ inside the FTI phase.
{\it Lower panel:} Phase diagram as a function of $U_{nnn}^{op}$ for spinless electrons. {\it Parameters for the exact diagonalization:} The noninteracting orbitals are determined on a $24 \times 24$ lattice with a pseudomagnetic flux $\Phi_0/48$ per hexagon (see supplementary material). The degeneracy of $n=0$ PLL is $N_s=12$ per spin direction and per valley. The low energy spectrum is calculated for $N_e=8$ ($N_L=N_R=4$) electrons with polarized spin occupying those $N_s=12$ states. Energies are given in units of $e^2/4 \pi \epsilon a_{0} \simeq 10$ eV.
}
\label{figPhaseDiagram}
\end{figure}

\bigskip

{\bf Fractionalized phases and superconductivity at $2/3$ filling of the $n=0$ PLL.}
Fractional Hall states in graphene under an external magnetic field were reported experimentally \cite{bolotin09,du09,dean2011}. Although strain produces flat PLLs, it is not evident that interactions can generate incompressible phases at fractional filling in time-reversal invariant strained graphene. We focus here on the $2/3$ filling of the four-fold degenerate $n=0$ PLL. This $2/3$ filling has been studied so far in graphene sheets \cite{toke2007,regnault2007} and in GaAs Hall bilayers \cite{mcdonald1996} under real magnetic field. In the present case of strained graphene, this particular filling allows for interesting possibilities including valley ferromagnetism (which breaks spontaneously time-reversal symmetry), valley symmetric topological states, and also superconductivity.

\begin{figure}
\vspace*{-2.0cm}
\hspace*{-2.0cm}
{\includegraphics[width=5.cm,height=6.8cm]{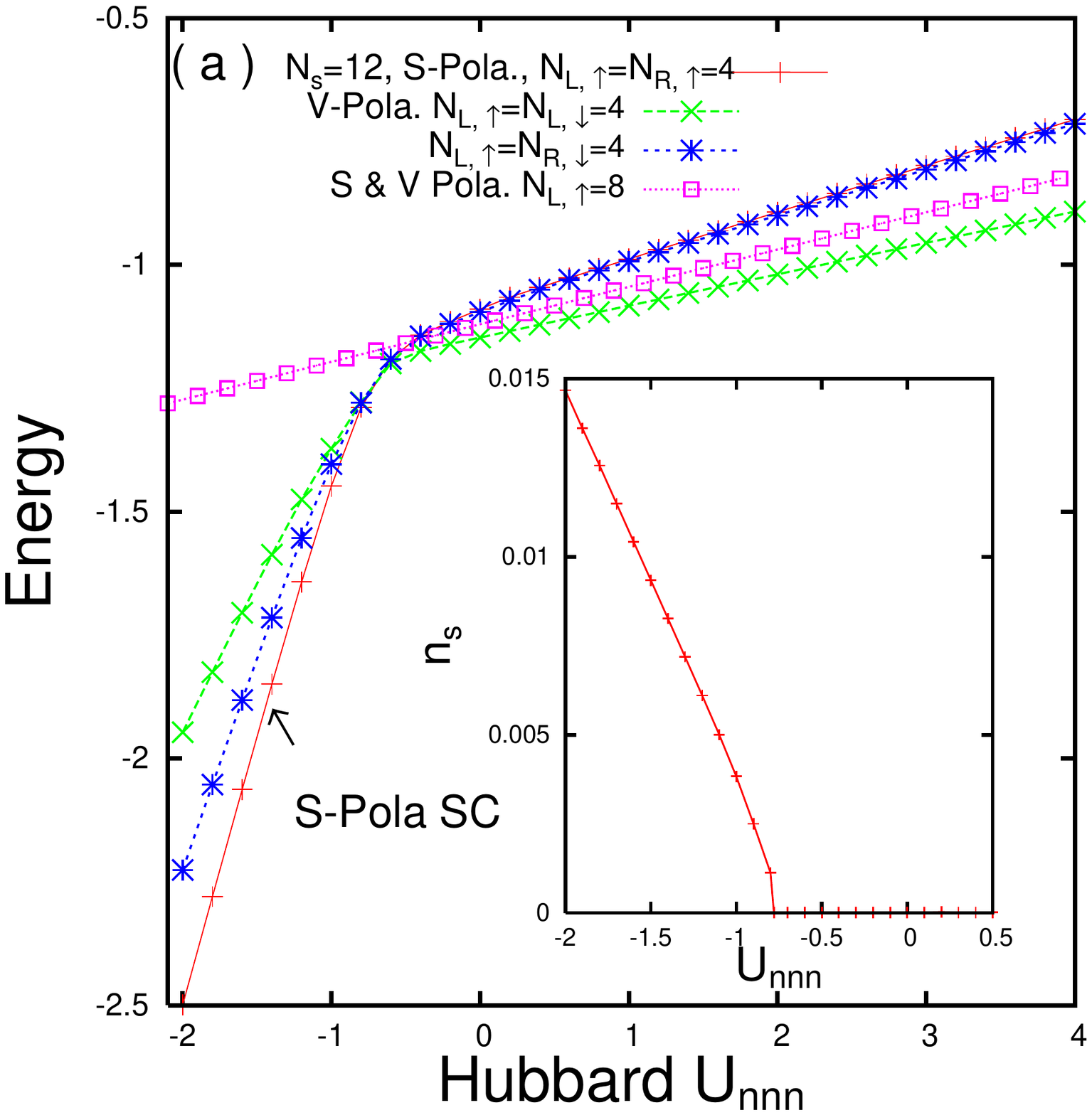}}
\hspace*{-0.5cm}
{\includegraphics[width=5.8cm, height=4.5cm]{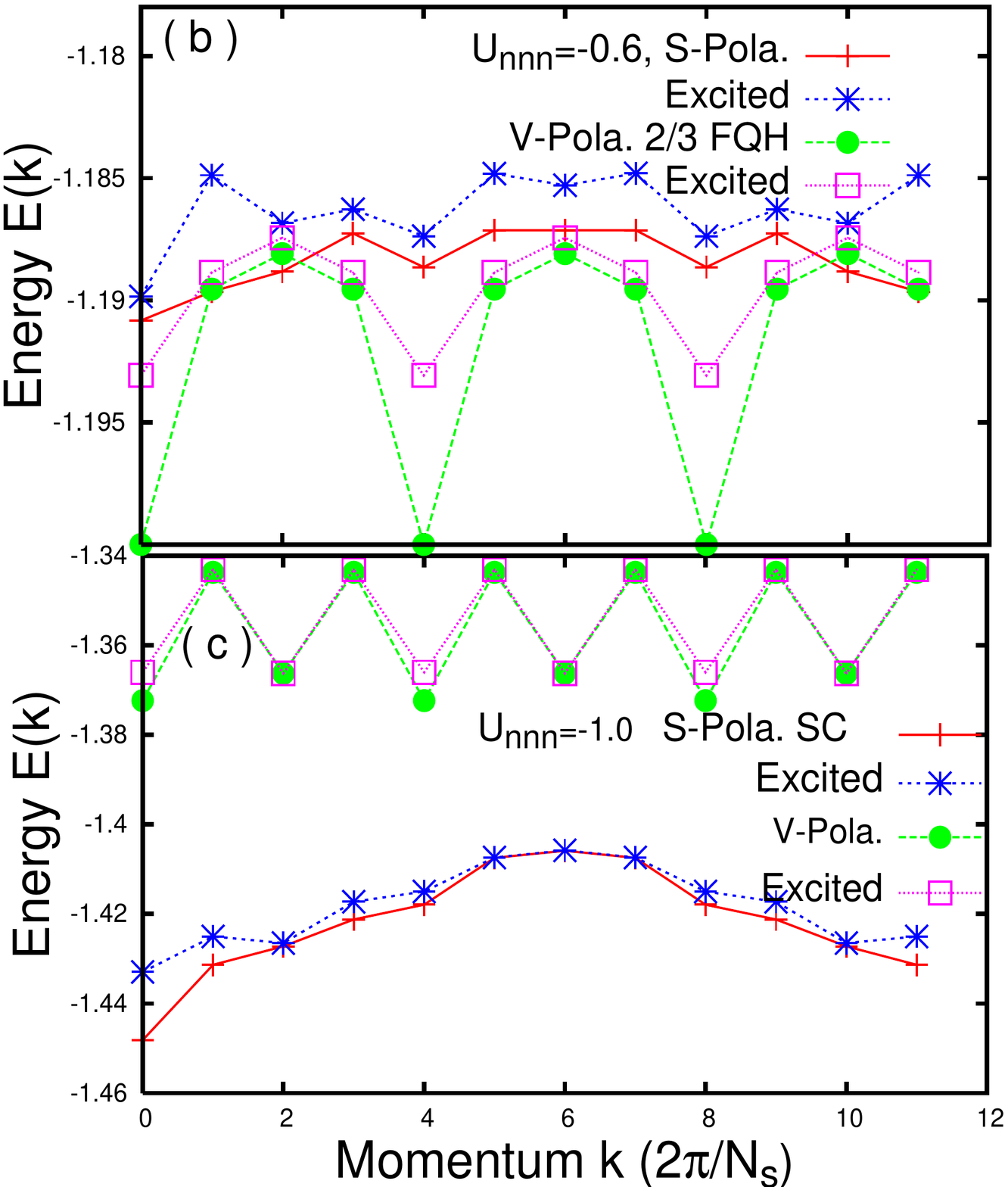}}
\hspace*{-2.0cm}
\vspace*{0mm}

\caption{(Color online) {\bf The $n=0$ PLL at fractional filling factor $\nu=-2+2/3$: spinfull electrons.} {\it Left panel:} The energy of different ground states as a function of the next-nearest-neighboring coupling $U_{nnn}$ defined in Eq. (\ref{hamiltonian2}). In the range $U_{nnn}<-0.8$, which includes the pure Coulomb interaction of realistic graphene ($U_{nnn}=0$), the ground state is a valley polarized and spin singlet FQH state (green crosses). Only a very significant attraction $U_{nnn}<-0.8$ can destabilize this state towards a valley unpolarized and spin polarized superconducting state (red crosses). This superconducting phase is characterized by a finite superfluid density as shown in the inset.
{\it Right panel:} Two lowest energies in each momentum sector as a function of $k$ for the valley-polarized state ($U_{nnn}=-0.6$, upper) and for the spin-polarized superconductor ($U_{nnn}=-1$, lower). {\it Parameters for the exact diagonalization:} same than in Fig. 1 but with the spin degree of freedom.}
\label{figNu23}
\end{figure}

{\it Real graphene: time reversal breaking FQH state in a single valley.} We first consider real graphene with the unscreened Coulomb interaction, namely $U_0=U_{nnn}=0$ in Eq. (\ref{hamiltonian2}). Then the ground state is found to be a valley polarized FQH state both for spinless (Fig. \ref{figPhaseDiagram}) and spinfull (Fig. \ref{figNu23}) electrons. This valley-polarized state breaks spontaneously the time-reversal invariance of the strained graphene Hamiltonian, and spins are in a singlet state as in the $2/3$ FQH states \cite{toke2007,regnault2007,mcdonald1996} obtained under real magnetic field. Due to the large values of strain-induced pseudomagnetic fields, this state may be realized with elevated energy scales, allowing for the stabilization of fragile states. In order to test quantitatively the robustness of the 2/3 valley polarized FQH state, we now vary the parameter $U_{nnn}$ in the Hamiltonian Eq. (\ref{hamiltonian2}). It turns out that the 2/3 valley polarized state is rather stable both in the spinless (Fig. \ref{figPhaseDiagram}) and spinfull (Fig. \ref{figNu23}) cases. Nevertheless when $U_{nnn}$ is sufficiently negative, exotic valley symmetric phases can also be realized as detailled below. For clarity we describe separately the spinless and spinfull cases.

{\it Spinless fermions and valley fractional topological insulator (FTI):} Let us consider spinless electrons and decompose the NNN coupling of Eq.(\ref{hamiltonian2}) into an interaction between opposite-valley electrons ($U_{nnn}^{op}$) and an interaction between same-valley electrons ($U_{nnn}^{s}$). We first tune the intervalley correlations ($U_{nnn}^{op}$) while $U_{nnn}^{s}=0$ (but note that electrons in the same valley still interact via the bare Coulomb potential).

In some intermediate parameter range ($-0.73<U_{nnn}^{op}<-0.58$), an interesting quantum phase emerges with nine nearly degenerated states forming a ground state manifold GSM (Fig. \ref{figPhaseDiagram}.a, lines with symbols) which is well separated from the higher energy states (Fig \ref{figPhaseDiagram}.a, lines without symbols). This valley-symmetric and 9-fold degenerated phase is called here valley fractional topological insulator, since valley plays the role taken by spin in the previously discussed "spin" FTIs \cite{levin2009,maciejko2010,Swingle2011,neupert2}. Moreover the momentum quantum numbers of these states are at $k=0$ and other $k$ determined by shifting the momentum of each electron by an integer multiple of $2\pi/N_s$, where $N_s$ is the PLL orbital degeneracy.  This determines three different momenta sectors ($k=0$, $\pi/3$, and $2\pi/3$) and there are three near degenerate states in each sector. These sectors can be considered as ground state flows from one sector to another upon inserting flux through adding the twist  boundary phase (Fig. \ref{figPhaseDiagram}.b). 

As a complementary characterization of the valley FTI phase, we further perform valley-pseudospin Chern number calculation \cite{thouless1982,niu1985} by adding the same boundary phase along $x$-direction, and the opposite ones along $y$-direction for both valleys \cite{balents2003,sheng2006}. This generalized pseudospin Chern number is well defined as the electron number in each valley is conserved thus that the valley-pseudospin is a good quantum number. We find a total Chern number quantized to 6 for all nine levels, characterizing the 2/3 fractionalized valley spin-Hall effect. 

Finally we can also turn on and increase the intravalley part of the NNN coupling $U_{nnn}^{s}$ (see supplementary). In the limit of very large intravalley correlations ($U_{nnn}^{s} \rightarrow \infty$), we expect a totally valley-decoupled  $1/3+1/{\bar{3}}$ phase consisting of two 1/3 Laughlin FQH states with opposite chiralities.  We find no phase transition between the valley FTI state discussed above (at  $U_{nnn}^{s}=0$ ) and the decoupled fractional valley Hall insulator (see supplementary).

The above results for spinless electrons can be summarized in a phase diagram (Fig. \ref{figPhaseDiagram}.c). For $U_{nnn}^{op}>-0.35$ (which includes realistic graphene), electrons have a natural tendency towards valley ferromagnetism, which is expected for repulsive interactions in a such a flat band system. In order to realize the valley FTI, one needs to counteract this trend by tuning an attraction between electrons in the opposite valleys. Besides, one also notices a narrow range of parameters ($-0.58<U_{nnn}^{op}<-0.35$) where the valley-polarization is lost but the GSM degeneracy not yet achieved. The understanding of this crossover region between the valley polarized FQH insulator and the valley FTI is still lacking and will be studied elsewhere. Finally superconductivity might appear when attraction is dominant ($U_{nnn}^{op}<-0.58$). This flat band superconductivity is discussed below in more details for the spinfull electrons.

{\it Spinfull fermions and spin triplet superconducting state.} We now consider spinfull fermions and we tune $U_{nnn}$ without distinguishing the valleys. For sufficiently large attraction (Fig. \ref{figNu23}), namely $U_{nnn} \leq -0.8$ (note that when added to the Coulomb repulsion, this ends up giving a somewhat smaller but still attractive next nearest neighbor interaction of $U^{\rm tot}_{nnn}=-0.2$), the ground state of the spinful model becomes a spin triplet and valley singlet superconducting state which is consistent with BCS-type mean field treatment (see supplementary). The superconductivity is characterized by a finite superfluid density $n_s=1/2\frac {\partial^2 E_g} {\partial \theta ^2}$  which is calculated from the change of the ground state energy $E_g$ upon adding a small phase twist $\theta$ as \cite{balents2003, ashvin2005}. Moreover the finite jump for $n_s$ at the transition point $U_{nnn}=-0.8$ (inset of Fig. \ref{figNu23}) points towards a first-order transition between the valley polarized state and the spin polarized superconducting state.  The typical momentum dependence of energy (Fig. \ref{figNu23}.c) differs drastically from the $2/3$ valley-polatized FQH case (Fig. \ref{figNu23}c) as the ground state is in the $k=0$-sector without the typical quasi-degeneracy of FQH state. 

\bigskip

{\bf Half-filling $n=0$ PLL.} We now turn to the case of neutral graphene (filling factor $\nu=0$) under large pseudomagnetic fields generated by strain. Due to the electron-hole symmetry, the $n=0$ PLL is half-filled and there is natural a competition between valley ferromagnet $\Psi_V=\prod_k c^{\dag}_{R,k,\uparrow} c^{\dag}_{R,k,\downarrow} | 0 \rangle$ and spin ferromagnet $\Psi_S=\prod_k c^{\dag}_{R,k,\uparrow} c^{\dag}_{L,k,\uparrow} | 0 \rangle$ ground states, $k$ labeling the Landau orbitals of the zero energy ($n=0$) PLL and ($R,L$) the valleys. Similar issue of valley and spin ferromagnetism in the half-filled $n=0$ Landau level has attracted a lot of interest for unstrained graphene under a real magnetic field \cite{alicea2006,goerbig2006,sheng2007,herbut2006,goerbig2011,kunyang}. Here we revisit this problem in the case of a time-reversal symmetric pseudomagnetic field.
\begin{figure}[h]
\vspace{-1.5cm}
\subfigure{\includegraphics[width=7.0cm, height=7.0cm]{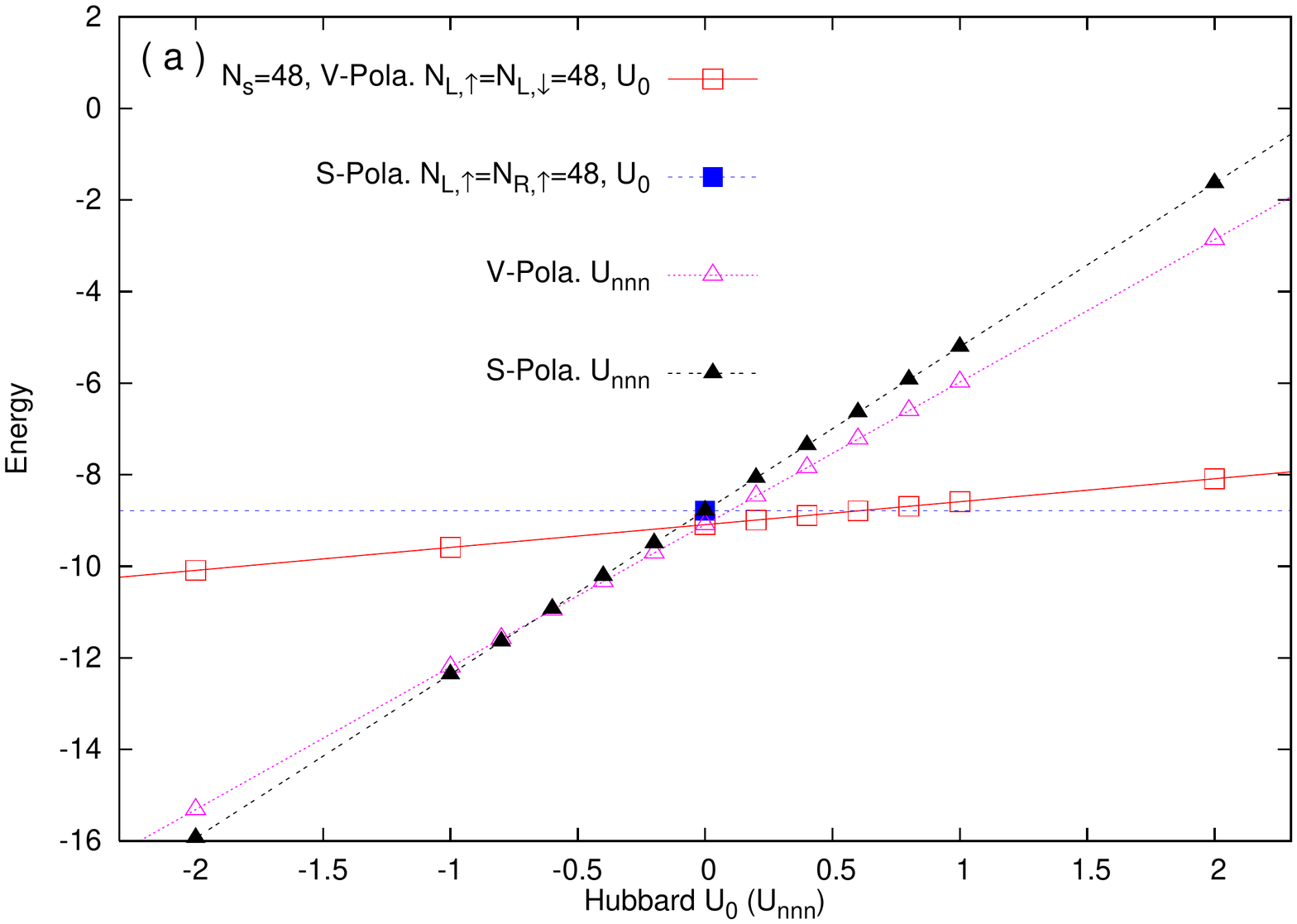}}
\vspace*{-1.7cm}
\subfigure{\includegraphics[width=7.0cm, height=4.5cm]{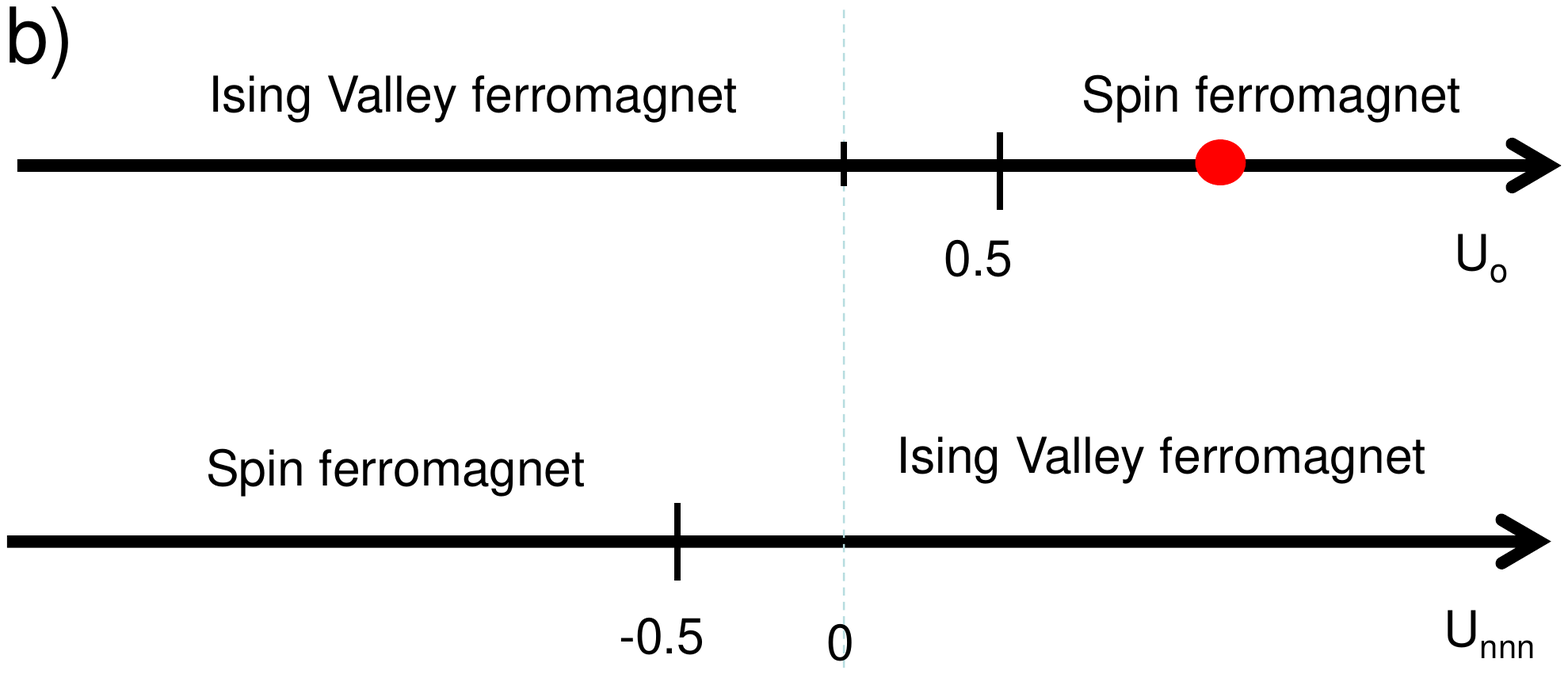}}
\caption{(Color online) {\bf Neutral graphene: $\nu=0$.} {\it Upper panel}: The Hartree-Fock energies of the Ising valley polarized ($\Psi_V$) and spin polarized ($\Phi_S$) states as a function of the on-site Hubbard coupling $U_0$ while $U_{nnn}=0$ (squares). We have also plotted the Hartree-Fock energies as a function of $U_{nnn}$ in the absence of on-site correlation, $U_0=0$ (triangles).  {\it Lower panel:} Corresponding phase diagrams. The energy is in units of $
e^2/4 \pi \epsilon a _{0}$ where $a_{0}$ is the distance between the nearest neighbor sites. The red dot indicates the value of the coupling $U_0$ for freestanding and unstrained graphene according to Ref. \cite{wehling2011}. {\it Numerical parameters:} The lattice we considered has $96\times 96$ sites. The degeneracy of the PLL orbits were $N_s=48$ while electron number is $N_e=96$. }
\label{figHalfFilling}
\end{figure}

We first consider the case of pure Coulomb interaction ($U_{0}=U_{nnn}=0$). Using Hartree-Fock method \cite{goerbig2006,alicea2006} we find that the valley and spin polarized states have the same energy when only dominant density-density terms are taken into account. We find that the intervalley backscattering terms lift this degeneracy by favoring the valley polarization. Note that for real magnetic field, those backscattering terms are absent in the zero-energy Landau level ($n=0$) due to the symmetry of the eigenspinors \cite{goerbig2006,alicea2006}. Also contrary to the real magnetic field Hall effect, long range Coulomb interaction prefers an Ising-like $Z_2$ valley polarized state rather than a more general $SU(2)$-valley-rotated state (see Supplemental material).

We now introduce {\it on-site Hubbard interaction} $U_0$ while $U_{nnn}=0$ in Eq. \ref{hamiltonian2} and compute numerically the total energy of finite size systems on a torus (Fig. \ref{figHalfFilling}, squares). As expected solely the energy of the valley polarized state is modified while the spin polarized state is unchanged (Fig. \ref{figHalfFilling}, horizontal dashed line) because double occupancy is forbidden by the Pauli principle in the fully spin polarized state. As a result, the competing valley polarized state (Fig. \ref{figHalfFilling}, empty squares) is the groundstate as long as the Hubbard interaction is not too repulsive ($U_0<0.5$ in units of $e^2/4 \pi \epsilon a_{0}$), including the pure Coulomb case. Further increase of the on-site Hubbard interaction stabilizes the spin ferromagnet state. Using gating or different substrates, it could be possible to switch the groundstate between spin ferromagnet and valley Ising ferromagnet. Spin polarized STM and Kerr imaging could indeed detect these competing ground states. Here the valley Ising ferromagnet is an integer quantum Hall state with two units of quantized Hall conductance, that spontaneously breaks time reversal symmetry.

In order to test the sensitivity of the phase diagram with respect to the details of the short range part of the interaction, we further consider a second model where the {\it next-nearest-neighbouring coupling} $U_{nnn}$ is varied while $U_0=0$. Interestingly we have obtained the reverse phenomenology where repulsive $U_{nnn}$ tends to valley polarize the system (Fig. \ref{figHalfFilling}, triangles).

{\bf Conclusion.} We have shown that strained graphene hosts various fractional topological phases which depend on the detailed structure of the
electron-electron interactions. In current experiments on both real graphene \cite{levy2010} and artificially designed molecular graphene \cite{gomes2011}, the nano-scale strained regions are small, typically of the order of the magnetic length, and they are strongly coupled to a metallic substrate. Future experiments on insulating substrates could address bigger strained regions. Nevertheless, signatures of fractional states in restricted domains and interactions with itinerant electrons outside the strained region will be important topics for future study. The $n=0$ Landau levels are expected to be the best isolated, since they occur at the Dirac point, where the density of itinerant states is the smallest. 

The predited phases relies on the flatness of the PLL $n=0$ which requires spacially homogeneous strained induced magnetic fields in each valley. To this respect artificially patterned honeycomb lattices \cite{West,gomes2011,Park} potentially allows for a better control upon the deformation pattern and therefore upon the flatness of the PLLs, in comparison to the mechanical strain in real graphene. Cold atoms in hexagonal optical lattices \cite{Sengstock,tarruell2011} are particularly suitable to access the attractive interaction regime. Finally we stress that the long range part of the Coulomb interaction is always present in our calculations. This is at odds with current experiments \cite{levy2010,gomes2011,Sengstock,tarruell2011} but it should be relevant for real graphene and for future experiments on artificial graphenes realised in surface states lying on insulating substrates.  Finally this study opens the prospect of discovering a series of new nontrivial topological phases at other fractional fillings and in higher pseudo Landau levels as well.

We are grateful to J. Alicea and N. Regnault for useful comments. JC acknowledges support from EU/FP7 under contract TEMSSOC and from ANR through project
2010-BLANC-041902 (ISOTOP).  This work is also supported by  DOE Office of Basic Energy Sciences under grant DE-FG02-06ER46305 (DNS), the NSF grant DMR-0958596 for instrument (DNS), and DE-AC02-05CH1123 (AV) as well as the Laboratory Directed Research and Development Program of Lawrence Berkeley National Laboratory under US Department of Energy Contract No.
DE-AC02-05CH11231 (PG). AV acknowledges hospitality from ICTS Bangalore where part of this work was completed. PG also acknowledges support from institute for condensed matter theory at University of Illinois at Urbana-Champaign.

Note: After submission of our preprint (arXiv:1111.3640), a study of non zero PLLs in strained graphene \cite{AbaninPreprint}, and related work on spin fractional topological insulators \cite{KunYangPreprint} appeared.

\bibliography{StrainGraphRefs}

\section{Supplementary}

\subsection{Noninteracting strained graphene}

Here we consider noninteracting spinless fermions on the honeycomb lattice (including of spin is straightforward). The triangular Bravais lattice $\bold{r}_{mn}=m \bold{a}_1 + n \bold{a}_2 $ is generated by the basis vectors:
\begin{equation}
\bold{a}_1 =\sqrt{3} a \bold{e}_x  \text{    and     }  \space \bold{a}_2 =\frac{\sqrt{3}a}{2} (\bold{e}_x + \sqrt{3} \bold{e}_y),
\label{basis}
\end{equation}
and the vectors:
\begin{equation}
\mbox{\boldmath$\delta$}_1 =\frac{a}{2}(\sqrt{3} \bold{e}_x + \bold{e}_y), \mbox{\boldmath$\delta$}_2  =\frac{a}{2}(-\sqrt{3} \bold{e}_x + \bold{e}_y) , \mbox{\boldmath$\delta$}_3 =- a \bold{e}_y,
\label{delta}
\end{equation}
connect any A atom to its three nearest B atoms, $a_{0}=0.142$ nm being the length of the carbon-carbon bond. The area of the unit cell is $\mathcal{A}_c =3 \sqrt{3} a_{0}^2 /2$.

\subsection{Strained induced gauge potential}

In the absence of interactions, the tight-binding Hamiltonian of strained graphene (Eq. 1 in the main text) can be written as:

\begin{equation}
\label{hamiltonian0rmn}
H_0 =\sum_{\bold{r}_{mn}} \sum_{a=1,2,3} (t+\delta t_{a} (\bold{r}_{mn})) ( a^\dagger (\bold{r}_{mn}) b (\bold{r}_{mn} + \mbox{\boldmath$\delta$}_a) + h.c. ),
\end{equation}
where second quantization operators $a(\bold{r}_{mn})$ and $b(\bold{r}_{mn} + \mbox{\boldmath$\delta$}_a)$  annihilate a fermion at A-type and B-type sites respectively. The strain is described by the deformation field $\delta t_{a}(\bold{r}_{mn})$ of the nearest-neighbour hopping element between sites $\bold{r}_{mn}$ and $\bold{r}_{mn} + \mbox{\boldmath$\delta$}_a$ with respect to the unperturbed value $t$. Note that the deviations $\delta t_{a}(\bold{r}_{mn})$ are real quantities and must be smaller than $t$.

In the absence of strain ($\delta t_{a}(\bold{r}_{mn})=0$), the lattice Hamiltonian can be easily diagonalized and the low energy excitations correspond to the states close to the two gapless Dirac points ${\bf K}_\xi=\xi (4 \pi/3 \sqrt{3}a_{0}) {\bf e_x}$, where $\xi=\pm$ is the valley isospin. Looking for low energy effective theory, we expand the annihilation operators as:
\begin{equation}
a (\bold{r}_{mn}) = a_{+} (\bold{r}_{mn}) e^{-i \bold{K} .\bold{r}_{mn}} + a_{-} (\bold{r}_{mn}) e^{i \bold{K} .\bold{r}_{mn}},
\label{enveloppe}
\end{equation}
in terms of the slowly-varying fields $ a_{\xi}(\bold{r}_{mn})$ (a similar equation holds for $b_\xi(\bold{r}_{mn})$ operators). Substituting into Eq. \ref{hamiltonian0rmn} and going from lattice to continuum fields as

\begin{equation}\label{continuum}\begin{split}
\sum_{\bold{r}_{mn}} & \rightarrow \int d^2 {\bf x}/\mathcal{A}_c   \text{             and             }  a_{\xi}(\bold{r}_{mn}) \rightarrow \sqrt{\mathcal{A}_c} a_{\xi}(\bold{x})  \text{           and           }  b_{\xi}(\bold{r}_{mn})\\ & \rightarrow \sqrt{\mathcal{A}_c} b_{\xi}(\bold{x}),
\end{split}\end{equation}
leads to the effective Hamiltonian:
\begin{equation}
H_0 =v_F \int d^2 {\bf x}  \sum_{\xi=\pm} \Psi_\xi^\dag({\bf x}) (\xi p_x^\xi \sigma_x + p_y^\xi \sigma_y) \Psi_{\xi}({\bf x}) ,
\label{hamiltoneffzerostrain}
\end{equation}
where $v_F=3at/2 \hbar$ is the Fermi velocity and $(p_x,p_y)=(-i \hbar \partial_x,-i \hbar \partial_y) $ are the components of the canonical momentum. The Pauli matrices $\sigma_x$ and $\sigma_y$ act on the lattice isospinors $\Psi_{\xi}({\bf x})=(a_\xi({\bf x}),b_\xi({\bf x}))$.

In the presence of a slowly varying deformation field $\delta t_{a}(\bold{r}_{mn})=\delta t_{a}(\bold{x})$, the valleys remain decoupled and the effective Hamiltonian reads:
\begin{equation}
H_0 =v_F \int d^2 {\bf x}  \sum_{\xi=\pm} \Psi_\xi^\dag({\bf x}) (\xi \Pi_x^\xi \sigma_x + \Pi_y^\xi \sigma_y) \Psi_{\xi}({\bf x}) ,
\label{effectivestrained}
\end{equation}
where $\mbox{\boldmath$\Pi$}^\xi= \textbf{p} + \xi e \textbf{A}$ shows the coupling of the electronic charge $-e$ with a valley-dependent gauge field $\textbf{A}_\xi (\textbf{x})=\xi \textbf{A}(\textbf{x})$ defined by:
\begin{equation}
e v_F (A_x(\textbf{x})+i A_y(\textbf{x}))=\sum_{a=1,2,3} \delta t_{a}(\textbf{x})e^{i \textbf{K}.\mbox{\boldmath$\delta$}_a} .
\label{hamilton1}
\end{equation}
More specifically using Eq. \ref{delta}, one obtains the gauge vector potential in terms of the deformation field:

\begin{eqnarray}
e v_F A_x &=& - \frac{1}{2} (\delta t_{1} + \delta t_{2} ) + \delta t_{3}, \nonumber\\
e v_F A_y &=& \frac{\sqrt{3}}{2} (\delta t_{1} - \delta t_{2} ),
\label{hamilton1}
\end{eqnarray}
where all the position arguments $\textbf{x}$ have been omitted.

\subsection{Example of a deformation pattern and its induced gauge field}
We now consider a particular deformation field where only the bonds along $\mbox{\boldmath$\delta$}_3 =- a \bold{e}_y$ are modified according to the pattern
\begin{equation}
(\delta t_{1},\delta t_{2},\delta t_{3})=e v_F B y(0,0,-1).
\end{equation}
According to Eq. 9, the corresponding strain-induced vector potential is the familiar vector potential of the Landau gauge $\bold{A} = -By \bf{e}_x$ describing here a uniform magnetic field $\bold{B}=B \bold{e}_z$ in the valley $\xi=+$, and the opposite field in valley $\xi=-$. In this gauge, the natural geometry is a rectangular one with linear sizes $L_x$ and $L_y$. Due to translational invariance, the system can be infinite in the $x$-direction but the smooth deformation condition ($\delta t_3 \ll t$) brings a limitation on the transverse size $L_y$ because $\delta t_3 $ is growing linearly long the $y$-direction. Assuming $\delta t_{3}(y=0)=0$, then $\delta t_{3}(L_y)=ev_F B L_y$ cannot exceed a reasonable fraction of $t=2 \hbar v_F/3a$ which leads to the condition:
\begin{equation}
L_y \ll \frac{\Phi_0}{\left | B \right |}\frac{1}{a}=\frac{l_B^2}{a}.
\end{equation}
Typically for a magnetic length $l_B=\sqrt{\hbar /e B} \simeq 10$ nm and lattice constant $a_{0} \simeq 0.1$ nm, the ribbon width cannot exceed $500$ nm. The strength of the effective magnetic field $B$ is proportional to the gradient of the hopping amplitude deformation. For a similar global deformation over the whole sample, a narrow ribbon hosts a stronger magnetic field than a broader ribbon.
Note that many other deformation fields lead to the same gauge vector potential,
including $(\delta t_{1},\delta t_{2},\delta t_{3})=e v_F B y(2,0,0)$ or $(\delta t_{1},\delta t_{2},\delta t_{3})=e v_F B y(0,2,0)$.

\subsection{Single electron wavefunctions}
We derive here the wavefunctions for noninteracting Dirac fermions under a
strong pseudo-magnetic field ($\xi B \bold{e}_z$), or more precisely the valley-dependent gauge potential $\textbf{A}_\xi (\textbf{x})=\xi \textbf{A}(\textbf{x})$, which are both opposite fields in the valleys $\xi=\pm$. We choose $B$ positive for definiteness, and denotes $l_B=\sqrt{\hbar /e B}$ the magnetic length. For each valley, we consider the first quantized Hamiltonian corresponding to Eq. \ref{effectivestrained}, namely:
\begin{equation}
h_\xi =v_F  \sum_\xi  (\xi \Pi_x^\xi \sigma_x + \Pi_y^\xi \sigma_y),
\label{hannex}
\end{equation}
where the components of the gauge-independent momentum $\mbox{\boldmath$\Pi$}^\xi= \textbf{p} + \xi e \textbf{A}$ do not commute due to the presence of the pseudo-magnetic field. Unlike the real magnetic field case, the sign of the commutator:
\begin{gather}
[\Pi_x^\xi,\Pi_y^\xi]=-i \xi \frac{\hbar^2}{l_B^2},
\label{commutator}
\end{gather}
depends on the valley index $\xi$. Hence the ladder operators are defined as
\begin{gather}
a_\xi=\frac{l_B}{\sqrt{2} \hbar}(\Pi_x^\xi -i \xi \Pi_y^\xi),
\label{ladder}
\end{gather}
in order to enforce the proper commutation relation $[a_\xi,a_\xi^\dag]=1$. The Hamiltonian can be written as:
\begin{equation}
h_\xi=\xi \frac{\hbar v_F \sqrt{2}}{l_B}%
\begin{pmatrix}
0 & a_\xi \\
a_\xi^\dag & 0
\end{pmatrix}%
.  \label{hmatrix}
\end{equation}%
We now focus on the zero energy Landau level. The corresponding wave function in the $\xi$-valley is $(0,v_\xi)$ with $a_\xi v_\xi =0$. Hence in the zero energy level, single electron wavefunctions are finite only on one triangular sublattice, here the B-atoms sublattice (since we have chosen the field strength $B$ to be positive). This is a general property valid for any strain-induced gauge field on the graphene lattice.

Now we give explicitly the wavefunctions for the Landau gauge field ${\bf{A}}=-By \bf{e}_x$. Then the equation $a_\xi v_\xi =0$ reads:
\begin{gather}
\left [ (-i \hbar \frac{\partial}{\partial x}-eB\xi y)-\xi \hbar \frac{\partial}{\partial y} \right] v_\xi(x,y)=0,
\label{groundstate}
\end{gather}
and substituting $v_\xi(x,y)=f_\xi(y)e^{ikx}$ in it, we get
\begin{equation}
 \left( \frac{d}{dy}+\frac{y-\xi k l_B^2}{l_B^2} \right)f_\xi (y)=0.
\label{eqdiff}
\end{equation}
Finally the normalized wavefunction is:
\begin{gather}
 v_\xi (x,y)=\frac{1}{\sqrt{\pi^{1/2} L_x l_B}} e^{- \frac{1}{2 l_B^2} (y-\xi k l_B^2)^2} e^{ikx} .
\label{eqdiff}
\end{gather}
Those wavefunctions correspond to the continuum model. For the numerical calculations performed on the honeycomb lattice and discussed in the main text, the wavefunctions are different from Eq. \ref{eqdiff} and have to be determined numerically using lattice model with pseudo magnetic field.
Nevertheless for the $n=0$-Landau level, there is a simple procedure to get the wavefunctions in the pseudo gauge field $\xi \bf{A}$ from the ones in the usual valley-independent gauge field $\bf{A}$. As said before for the real magnetic field, the $(\xi=+)$-valley wavefunctions have predominant weight on B atoms, and $(\xi=-)$-valley wavefunctions on A atoms. To get the corresponding wavefunctions in a pseudo magnetic field, one has simply to i) keep the former $(\xi=+)$-valley wavefunctions on B atoms without any change, ii) remove the wavefunctions on A sites and finally iii) add complex conjugates of B sites wavefunctions.

\subsection{Effect of interactions}

Here we will consider the projection of density-density interactions of the form $H_I=\sum_{\textbf{r},\textbf{r'}} V(\textbf{r}-\textbf{r'})n(\textbf{r})n(\textbf{r'})$ into the $n=0$ PLL which then would have the form:

\begin{equation}
H^{n=0}_I=\sum_{P_1,P_2,P_3,P_4,\sigma,\sigma'} V_{P_1,P_2,P_3,P_4}\  c^\dagger_{\sigma,P_1} c^\dagger_{\sigma',P_2} c_{\sigma',P_3} c_{\sigma,P_4},
\end{equation}
where $c^\dagger_{P,\sigma}$ is the creation operator of fermion in state $P=\{\xi,k\}$ which is in valley $\xi$ and Landau orbital $k$ (see eqn. \ref{eqdiff}).

Using the continuum model derived in the last section we get the the projected potential:

\begin{equation}\label{project}\begin{split}
& V_{P_1,P_2,P_3,P_4} =\left(\frac{g^2 a^2}{L_x\  l}\right)^2 \int d\textbf{r} d\textbf{r}' V(\textbf{r}-\textbf{r}') (-i)^{\xi_3+\xi_4-\xi_1-\xi_2} \\ & e^{-\frac{1}{4 l^2}\left[ (k_1 l^2-y)^2+(k_2 l^2-y')^2+(k_3 l^2-y')^2+(k_4 l^2-y)^2 \right]} \\
& e^{i\frac{4\pi}{3a} \left[ (\xi_1-\xi_4)x+(\xi_2-\xi_3)x' \right]}\ e^{-i\left[(k_1\ \xi_1-k_4\ \xi_4)x+(k_2\ \xi_2-k_3\ \xi_3)x'\right]},
\end{split}\end{equation}
where $g=\left(\frac{3}{64\pi}\right)^{\frac{1}{4}}$. We can now consider different interaction potentials.

If the interaction is smooth the dominant interaction term has $\xi_1=\xi_4$ and $\xi_2=\xi_3$.

There are two polarized states which will be particularly of our interest. valley  $|\psi_V\rangle =\Pi_k c^\dagger_{1,k,\uparrow}c^\dagger_{1,k,\downarrow}$ or spin $|\psi_S\rangle =\Pi_k c^\dagger_{1,k,\uparrow}c^\dagger_{-1,k,\uparrow}$ polarized states.  The energy of these states have the general form of:

\begin{equation}\label{valleypl}\begin{split}
E_{V} =\sum_{\{p_1,1\},\{p_2,1\},\sigma,\sigma'} & V_{\{p_1,1\},\{p_2,1\},\{p_2,1\},\{p_1,1\}}\\ & -V_{\{p_1,1\},\{p_2,1\},\{p_1,1\},\{p_2,1\}}\delta_{\sigma,\sigma'},
\end{split}\end{equation}
and
\begin{equation}
\label{spinpl}\begin{split} E_{S} =\sum_{\{p_1,\xi_1\},\{p_2,\xi_2\}} & V_{\{p_1,\xi_1\},\{p_2,\xi_2\},\{p_2,\xi_2\},\{p_1,\xi_1\}}\\ &-V_{\{p_1,\xi_1\},\{p_2,\xi_2\},\{p_1,\xi_1\},\{p_2,\xi_2\}}\delta_{\xi_1,\xi_2}.
\end{split}
\end{equation}

\subsection{Coulomb interaction}

Putting the Coulomb interaction $V({\bf r}_i-{\bf r}_j)n({\bf r}_i)n({\bf r}_j)=e^2n({\bf r}_i)n({\bf r}_j)/4 \pi \epsilon|\bf r_i - \bf r_j|$ in projection \ref{project}:

\begin{equation}\begin{split}
V_{P_1,P_2,P_3,P_4} =& \left(\frac{g^2 a^2}{L_x\  l}\right)^2 \int  d\textbf{r} d\textbf{r}' \frac{q^2}{|\textbf{r}-\textbf{r}'|} \\ & e^{-\frac{1}{2 4l^2}\left[ (k_1 l^2-y)^2+(k_2 l^2-y')^2+(k_3 l^2-y')^2+(k_4 l^2-y)^2 \right]} \\ &   e^{-i\left[\xi_1(k_1-k_4)x+\xi_2(k_2-k_3)x'\right]}
\end{split}\end{equation}

For neutral graphene the $n=0$ PLL is at half filling. The interactions naturally prefers the polarized state of valley or spin degrees of freedom. Using \ref{spinpl} the energy of spin polarized state is given by:

\begin{equation}\begin{split}
E_{S} =\sum_{p_1,p_2} & \left(\frac{g^2 a^2}{L_x\  l}\right)^2 L_x
 \int d(x-x')\ \ dy dy' \frac{q^2}{\sqrt{(x-x')^2+(y-y')^2}} \\  & 2\left(2e^{-\frac{1}{2 l^2}\left[ (p_1 l^2-y)^2+(p_2 l^2-y')^2 \right]} - \cos{\left[(p_1-p_2)(x-x')\right]}\right. \\ & \left. e^{-\frac{1}{4 l^2}\left[ (p_1 l^2-y)^2+(p_1 l^2-y')^2+(p_2 l^2-y')^2+(p_2 l^2-y)^2 \right]}\right)
\end{split}\end{equation}

Interestingly the valley polarized state has the same energy: $E_V=E_S$.   If we include the fast oscillation which we ignored before ($e^{i\frac{4\pi}{3a} \left[ (\xi_1-\xi_4)x+(\xi_2-\xi_3)x' \right]}$), the valley polarized state does not change since $\xi_1=\xi_2=\xi_3=\xi_4$, but the energy of spin polarized state increases, so the valley polarized state will be the ground state of neutral graphene if the long range Coulomb is the dominating interaction.

Notice that we considered a valley polarized state in a single valley. If we consider an arbitrary rotated valley polarized state in a state of the form:

\begin{equation}
|\psi_{V'}\rangle =\Pi(u^+_k\ c^\dagger_{+,k,\uparrow}+u^-_k\ c^\dagger{-,k,\uparrow})(u^+_k\ c^\dagger_{+,k,\downarrow}+u^-_k\ c^\dagger_{-,k,\downarrow})|0\rangle ,
\end{equation}
the Hartree-Fock energy have the form:

\begin{equation}
E_{V'}=E_V+2\sum_{p_1\neq p_2} \Gamma_{p_1,p_2}\sum_\xi|u^{-\xi}_{p_1}|^2|u^{\xi}_{p_2}|^2 ,
\end{equation}
where $\Gamma_{p_1,p_2}$ is positive for the Coulomb potential. The term added is then always positive and the minimum of energy is for
$u^{+}_{p}=0$ or $u^{-}_{p}=0$ for all $p$. The the preferred state would be an Ising valley polarized state.

\subsection{Short range Hubbard interactions and superconductivity}

Although all types of density-density interactions could be treated using our projection scheme, here we only present the details for the on-site and next nearest neighbor interactions which we also studied numerically and observed the somehow unexpected spin polarized superconductivity.

The on-site interaction has the form $\int d\textbf{r} d\textbf{r}' V \delta(\textbf{r}-\textbf{r}')$ where as next-nearest neighbor interaction has the form $\int d\textbf{r} d\textbf{r}' V\sum^6_{i=1}\delta(\textbf{r}+\textbf{R}_i-\textbf{r}')$ where $\textbf{R}_i=(X_i,Y_i)$ are the vectors connecting each site to its six next nearest neighbors. This next nearest neighbour interaction projected into the $n=0$ PLL has the form:

\begin{equation}\begin{split}
V_{P_1,P_2,P_3,P_4}= & V\left(\frac{g^2 a^2}{L_x\  l}\right)^2 L_x \int dy\\  & e^{-\frac{1}{ l^2}\left[ (y-\frac{p_1+p_2+p_3+p_4}{4}l^2)^2 +l^4\frac{\sum(p_i-p_j)^2}{32}\right]} \Omega(\xi_2-\xi_3).
\end{split}\end{equation}
Here we have $\xi_2-\xi_3=\xi_4-\xi_1$ and $\xi_1 p_1+\xi_2 p_2-\xi_3 p_3-\xi_4 p_4=0$. Notice that the interaction is not smooth in the lattice scale so we should keep the oscillatory term $\Omega(\xi_2-\xi_3)=\sum^6_{i=1}e^{i\frac{4\pi}{3 a}(\xi_2-\xi_3)X_i}=2\cos{\left[\frac{4\pi}{3}  (\xi_2-\xi_3) \right]}+4\cos{\left[\frac{2\pi}{3}  (\xi_2-\xi_3) \right]}$.  %$U(p_1,p_2,p_3,p_4)$ is invariant under any permutation of the momentums $p_1,p_2,p_3$ and $p_4$.

For on-site interaction we have similar form with $\Omega(\xi_2-\xi_3)=1$.

%With this projected interaction one can readily check that for next nearest neighbour repulsive interaction, valley polarized state is preferred, where is attractive interaction spin polarized state have lower energy.

%With the projected interaction of the general form  $\sum_{P_1,P_2,P_3,P_4,\sigma,\sigma'} V_{P_1,P_2,P_3,P_4}\  c^\dagger_{\sigma,P_1} c^\dagger_{\sigma',P_2} c_{\sigma',P_3} c_{\sigma,P_4}$ we can also compare the mean-field energy of different superconducting states. We compare the energy of two superconducting states which are valley triplet,spin singlet:

With this form of the projected interaction potential we can readily compare the energy of spin and valley polarized states. Using the expressions in \ref{valleypl} and \ref{spinpl} we get:

\begin{equation}\begin{split}
E_{V}= & V\frac{g^4 a^4}{L_x\  l^2}\int dy\ \sum_{p_1,p_2}e^{-\frac{1}{ l^2}\left[ (y-\frac{p_1+p_2}{2}l^2)^2 +l^4 \frac{(p_1-p_2)^2}{16}\right]}\\ & \sum_{\sigma,\sigma'}\Omega(0)(1-\delta_{\sigma,\sigma'}),
\end{split}\end{equation}
and
\begin{equation}\begin{split}
E_{S}= & V\frac{g^4 a^4}{L_x\  l^2}\int dy\  \sum_{p_1,p_2} e^{-\frac{1}{ l^2}\left[ (y-\frac{p_1+p_2}{2}l^2)^2 +l^4 \frac{(p_1-p_2)^2}{16}\right]} \\ & \sum_{\xi_1,\xi_2}\Omega(0)-\Omega(\xi_2-\xi_1).
\end{split}\end{equation}

For the on-site interaction with $\Omega(\xi_2-\xi_1)=1$  the spin polarized state has no energy-gain where as valley polarized state energy increases with on site repulsion.

For the next nearest neighbour interaction $\Omega(\xi_2-\xi_1)=\Omega(0)(\frac{3}{2}\delta_{\xi_1,\xi_2}-1)< \delta_{\xi_1,\xi_2}$. So valley  polarized state will be stabilized with next nearest neighbor interactions.

With the projected interaction of the general form  $\sum_{P_1,P_2,P_3,P_4,\sigma,\sigma'} V_{P_1,P_2,P_3,P_4}\  c^\dagger_{\sigma,P_1} c^\dagger_{\sigma',P_2} c_{\sigma',P_3} c_{\sigma,P_4}$ we can also compare the mean-field energy of different superconducting states. We compare the energy of two superconducting states which are valley triplet,spin singlet:

\begin{equation}
|\Psi_{Singlet} \rangle=\Pi_{P_i=\{+,p_i\},\{-,p_i\}}(u_i+v_i c^\dagger_{P_i,\uparrow}c^\dagger_{\bar{P}_i,\downarrow})|0\rangle ,
\end{equation}
and spin triplet, valley singlet:

\begin{equation}
|\Psi_{Triplet} \rangle= \Pi_{P_i=\{+,p_i\},\sigma=\uparrow,\downarrow}(u_i+v_i c^\dagger_{P_i,\sigma}c^\dagger_{\bar{P}_i\bar{\sigma}})|0\rangle ,
\end{equation}
where $\bar{P}$ and $\bar{\sigma}$ are  the time reversal of $P$ and $\sigma$ respectively.

The corresponding mean-field equations for the superconducting gap for the spin singlet state reads:

\begin{equation}
\Delta_P=-\sum_{P'=\{+,p_i\},\{-,p_i\}} V_{P,\bar{P},\bar{P'},P'} u_{P'}v_{P'} ,
\end{equation}
and for the spin triplet state reads:

\begin{equation}
\Delta_P=-\sum_{P'=\{+,p_i\},\sigma=\uparrow,\downarrow} V_{P,\bar{P},\bar{P'},P'} u_{P'}v_{P'} ,
\end{equation}
where

\begin{equation}\nonumber \begin{split}
V_{P,\bar{P},\bar{P'},P'}=& V\left(\frac{g^2 a^2}{L_x\  l}\right)^2 L_x \int dy\ e^{-\frac{1}{ l^2}\left[ (y-\frac{p+p'}{2}l^2)^2 +l^4\frac{\sum(p-p')^2}{16}\right]} \\ & \left(2\cos{\left[\frac{4\pi}{3}  (\xi_2-\xi_3) \right]}+4\cos{\left[\frac{2\pi}{3}  (\xi_2-\xi_3) \right]}\right)
\end{split}\end{equation}

In the spin-singlet state $V_{P,\bar{P},\bar{P'},P'}$ changes sign between the states with $\xi'=-\xi$ and $\xi'-=\xi$, where as for the spin triplet for all the term in the mean field equation have $\xi'=\xi$. So we get larger gap and so smaller mean-field energy for the spin triplet superconducting state.

\subsection{The robustness of the fractional valley Hall insulator and its evolution
with tunning $U_{nnn}^s$}

\begin{figure}[htp]
%\vspace*{-10mm}
{\includegraphics[width=9.cm,height=7.5cm]{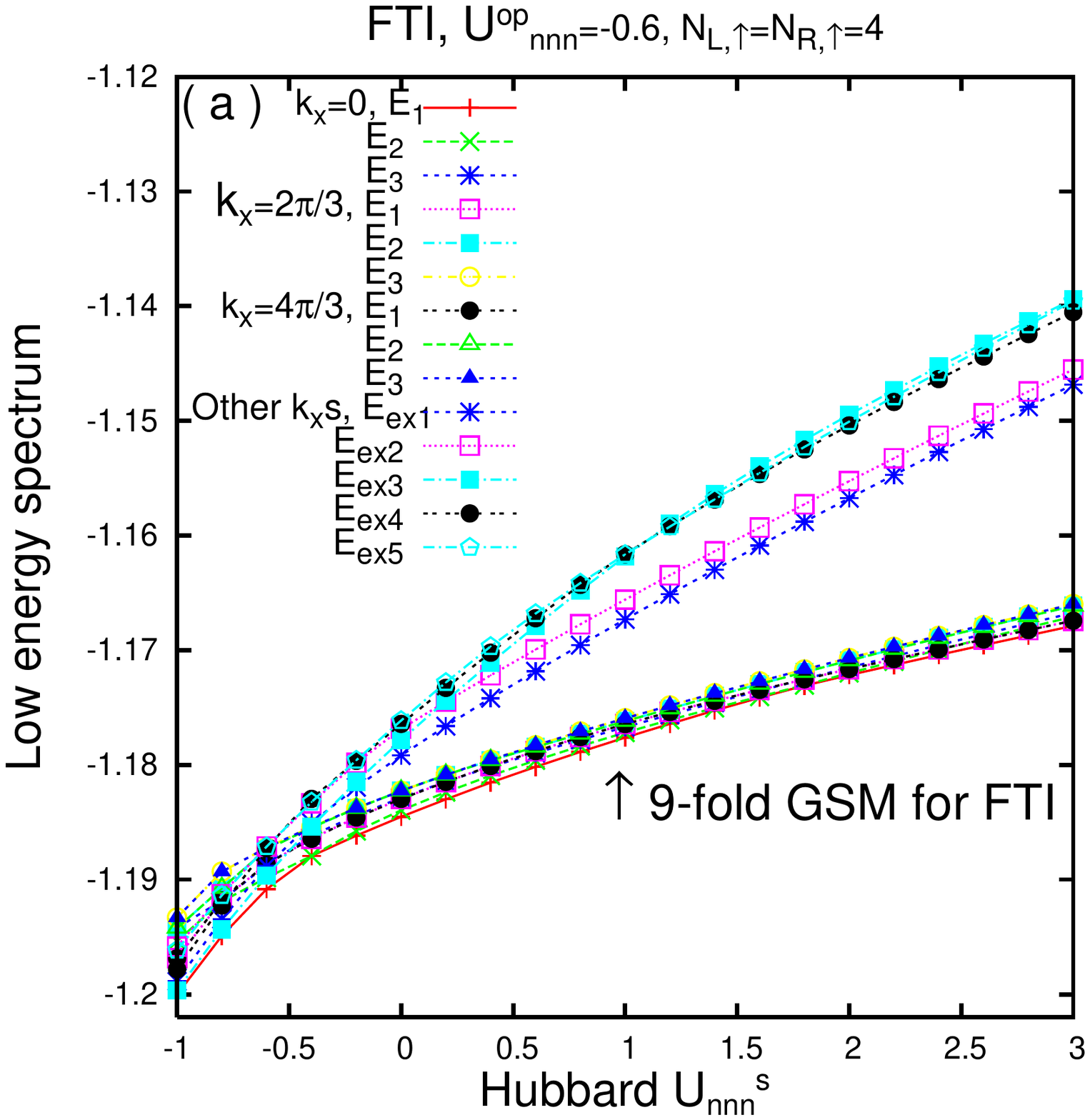}}
{\includegraphics[width=9.cm,height=7.5cm]{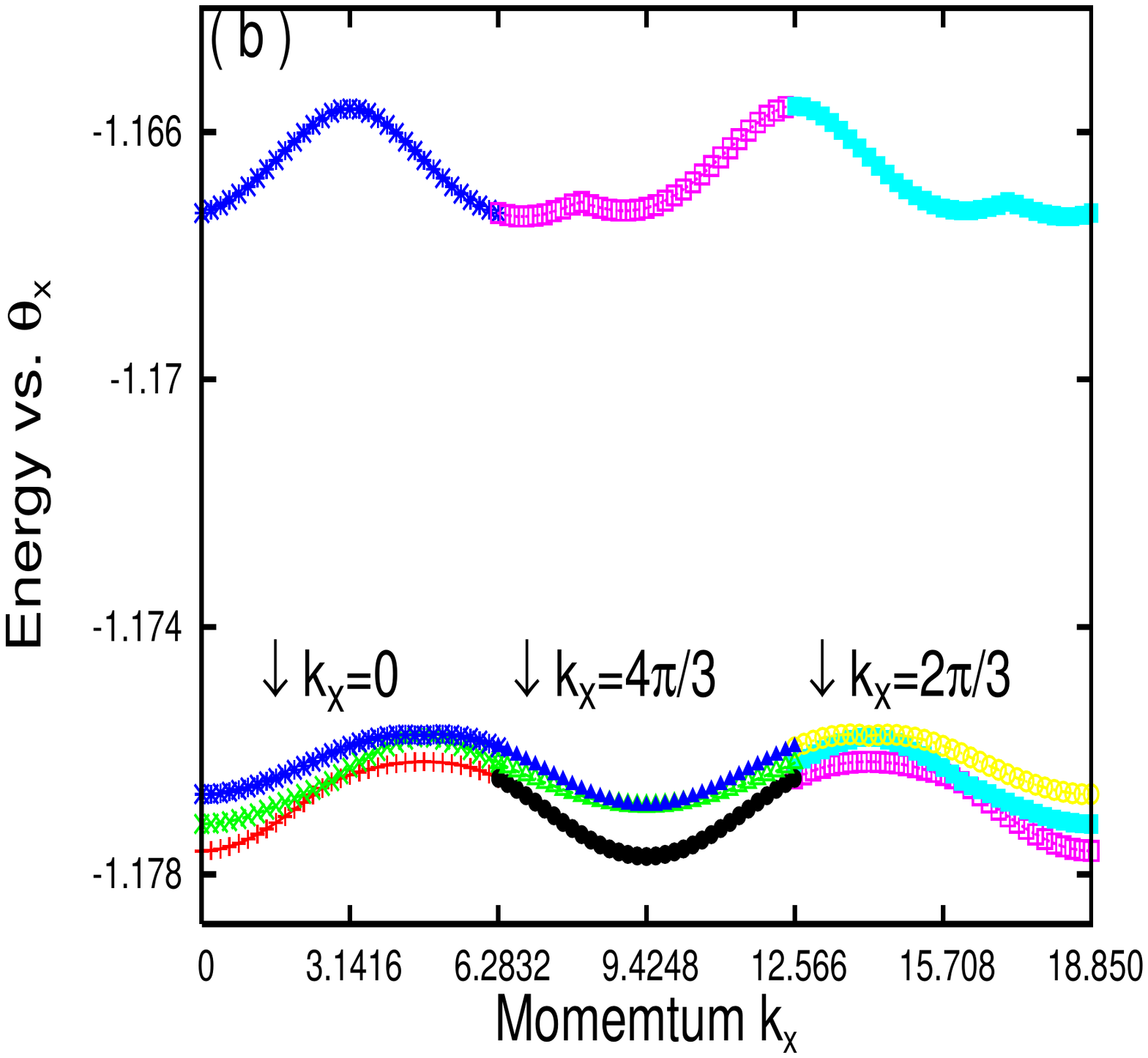}}
\caption{(Color online) (a) Fractional topological insulating phase for spinless electrons at filling $\nu=-2+2/3$.
The LL degeneracy is $N_s=12$ while there are totally
 $N_e=8$ ($N_L=N_R=4$) electrons with polarized spin on a lattice with $24*24$ sites.
We demonstrate an emergent symmetry where energies of nine states
from  $k=0$, $2\pi/3$ and $4\pi/3$ sectors become near degeneracy at
large $U_{nnn}^s$ limit.  The onset of the fractionalized phase with
 nine-fold near degenerating GSM  is identified
at $U_{nnn}^s\geq -0.2$. (b) The evolution and the robust of the gap
during the change of the boundary phase.
}
\label{fig5}
\end{figure}

The fractional valley Hall insulator is characterized by a large ground state
degeneracy.  In a finite-size system, due to the coupling between different
states, one usually sees a quasi-degeneracy with a finite splitting
between these states in the ground state manifold.
As shown the main text,  around $U_{nnn}^{op}=-0.6$ without the interaction between the electrons in the same valley $U_{nnn}^s=0$, the nine states at the right quantum number sectors are indeed have much lower energy than other excited states.
However, the splitting between these states are close to the finite gap between these states and other excited states. While the quantized nonzero total Chern number of the ground state manifold
indicates the obtained state is indeed a
fractional valley Hall insulator, a clear-cut evidence of nine fold
topological degeneracy
is still absent and it is difficult to predict the fate of the state
as system becomes very large.
Here we address this issue  through tuning the system deep into
the topological phase. Indeed this can be achieved by increasing the
correlations between the electrons in the same valley ($U_{nnn}^s$).

As shown in Fig. \ref{fig5}, with the turn on of  positive  $U_{nnn}^{s}$,
the energy gap between the ground state manifold
and other  excited states becomes very robust and much larger than the
splitting of the energy of the ground state manifold.
There is no phase transition as $U_{nnn}$ continuously increases,
 so the observed state is indeed the
same phase as the  fractional valley Hall insulator at the decoupled limit
 (strong $U_{nnn}^s$ limit). In that limit,  the nine fold degeneracy is exact
and spinless electrons in different valleys are
 contributing $\pm 1/3$ quantized Hall conductances.
We further perform the flux inserting measurement. we show in Fig. \ref{fig5}b
for $U_{nnn}^s=1$,
the three lowest energy states in the momentum sector $k=0$ evolve into other
states in the ground state manifold and they evolve back to themselves after
three periods of boundary phase insertion.  The energy gap between these states and
other excited states remain robust as illustrated in Fig. \ref{fig5}b. We further
perform valley-dependent Chern number calculation,
%by adding the same boundary phase along x-direction, and the opposite ones
%along y-direction for both valleys\cite{balents2003, sheng2005}. This
%generalized pseudospin Chern number is well defined as the electron number in
%each valley is conserved.
and find a total Chern number quantized to 6 for all
nine levels, characterizing the 2/3 fractionalized valley spin-Hall
effect.
Remarkably, this phase
persists in a wide range of $U_{nnn}^s \geq -0.2$ including the
simple case where this interaction is turned off ($U_{nnn}^s=0$
as shown in the main text part of the paper.

\end{document}